\documentclass[a4paper]{PoS}
\usepackage[utf8]{inputenc}
\usepackage[T1]{fontenc}
\pdfoutput=1
\usepackage{amsmath}
\usepackage{marvosym}
\usepackage{color}

\newcommand{\p}[1]{\left( #1 \right)} 
\newcommand{\s}[1]{\left[ #1 \right]} 

\newcommand{\phys}{\text{phys}}

\title{
  \begin{picture}(0,0)(0,0)%
    \put(300,75){\makebox(0,0)[l]{\textnormal
        {\normalsize KEK-CP-351, OU-HET-921 }
      }}
  \end{picture}Approaching the Bottom Using Fine Lattices With Domain-Wall Fermions
}

\ShortTitle{Approaching the Bottom Using Fine Lattices With Domain-Wall Fermions}

\author{\speaker{Brendan Fahy}$^a$, Guido Cossu$^a$, Shoji Hashimoto$^{ab}$\\
  {$^a$}High Energy Accelerator Research Organization (KEK), Ibaraki 305-0801, Japan\\
  {$^{b}$}School of High  Energy Accelerator Science, SOKENDAI (The Graduate University for
  Advanced Studies), Tsukuba 305-0801, Japan}

\abstract{ We explore the heavy-quark mass region above the charm mass
  using M\"obius domain-wall fermions on fine lattices at $a = 0.080$, $0.055$, and $0.044$ fm. We examine masses and decay
  constants using a series of heavy quark masses up to 3 times the
  charm quark. We analyze the cutoff effects for heavy quarks above
  the charm and account for the leading order discretization effects
  using ideas from HQET.  We extrapolate to the bottom quark mass and
  report preliminary results for $f_{B}$ and $f_{B_s}$.}

\FullConference{34th annual International Symposium on Lattice Field Theory\\
  24-30 July 2016\\
  University of Southampton, UK}

\begin{document}

\section{Introduction and Lattice Setup}

Direct lattice simulation of the bottom quark is still a challenge for
lattice QCD.  One approach is to use effective actions specific for
heavy quark physics such as non-relativistic QCD and match back to
QCD.  Modern lattices are, on the other hand, much finer than in the
past and cutoff effects at the charm are small. Using very fine lattices,
cutoff effects are manageable even above the charm
mass~\cite{Fahycharm}, which allows us to produce results between the
charm and the bottom with enough points to extrapolate to the bottom.

The JLQCD collaboration has recently produced lattice ensembles with $2+1$ flavors of
M\"obius domain-wall fermions~\cite{mobius}. The
gauge action used is tree-level improved Symanzik. These lattices have
relatively fine lattice spacings of $1/a \approx 2.4$, $3.6$,
and $4.5\text{ GeV}$ with pion masses between $230$ MeV and $500$
MeV~\cite{finelattice}. The lattice spacing was determined from Wilson
flow using $t_0$ with the physical value from \cite{scale}.
The parameters of each of the $15$ lattices can be found in
Table~\ref{tab:lattices}.

Axial and pseudo-scalar two-point correlators were computed with the
Iroiro++ software package~\cite{iroiro}. Correlators were improved
using $Z_2\,(\pm1)$ noise sources distributed over a single time
slice, and sources computed on many time slices on a single
configuration are then averaged. A total of $400-600$ measurements
were carried out as detailed in Table~\ref{tab:lattices}. Each of
these $Z_2$ sources were computed both unsmeared and with Gaussian
smearing.  The source-sink combinations of unsmeared-smeared and
smeared-smeared were simultaneously fit to extract meson masses and
decay constants. Due to good chiral symmetry of our domain-wall
fermions the decay constants can be computed directly from the
psuedo-scalar currents utilizing the PCAC relation. These lattices
have also been used to compute semi-leptonic $D$ decays in
\cite{Kaneko2017sct} and charm quark mass determination from
short-distance correlators in \cite{PhysRevD.94.054507}.

\begin{table}[tbph]
  \centering
  \begin{tabular}{|l|l|r|r|r|r|r|r|}
    \hline
    $\beta$  &                         $L^3\times T$  &   $L_5$  &$a m_{ud}$  &$a m_s$  &$  m_\pi $  &$ m_{\pi}L $   &\#meas \\
      &                                &                  &       &            &         [MeV]  &     &               \\
    \hline
    $\beta = 4.17$  &$ 32^3\times64$  & 12  &   0.0035  &    0.040  &  230  &       3.0         &   800  \\
      &    &                  &       0.0070  &    0.030  &  310  &       4.0          &  800 \\
             &                         &                  &       0.0070  &    0.040  &  310  &       4.0           & 800\\
             &                         &                  &       0.0120  &    0.030  &  400  &       5.2        &    800   \\
             &                         &                  &       0.0120  &    0.040  &  400  &       5.2     &       800      \\
             &                         &                  &       0.0190  &    0.030  &  500  &       6.5  &          800         \\
             &                         &                  &       0.0190  &    0.040  &  500  &       6.5      &      800     \\ \cline{2-7}
             &                         $48^3\times96 $  & 12  &   0.0035  &    0.040  &  230  &       4.4     &       800      \\
    \hline
    $\beta= 4.35$  & $48^3\times 96 $  &8  &    0.0042  &    0.018  &  300  &       3.9      &      600     \\
      & &                  &       0.0042  &    0.025  &  300  &       3.9        &    600   \\
             &                         &                  &       0.0080  &    0.018  &  410  &       5.4        &    600   \\
             &                         &                  &       0.0080  &    0.025  &  410  &       5.4      &      600     \\
             &                         &                  &       0.0120  &    0.018  &  500  &       6.6    &        600       \\
             &                         &                  &       0.0120  &    0.025  &  500  &       6.6  &          600         \\
    \hline
    $\beta = 4.47$  &$64^3\times128 $  &8  &    0.0030  &    0.015  &  280  &       4.0      &      400     \\
    \hline
  \end{tabular}
  \caption{Parameters of the JLQCD gauge ensembles used in this
    work. Pion masses are rounded to the nearest $10$ MeV. Inverse
    lattice spacings are $a^{-1}=2.453(4)\text{ GeV}$,
    $3.610(9)\text{ GeV} $, and $4.496(9) \text{ GeV}$ for
    $\beta=4.17,4.35$ and $4.47$ respectively. The $L_5$ length in the
    domain-wall is $12$ at $\beta=4.17$ and $8$ on the finer
    lattices. The spatial extent satisfies $m_\pi L \gtrsim \geq 4.0$
    for all lattices except the first lattice in the table, which was
    excluded from the final analysis in this work. The number of
    measurements listed in the final column are the product of the
    number of configurations used and the number of time sources on
    each configurations. $100$ configuration's were used for the
    $\beta=4.17$ ensembles and $50$ in the
    others. \label{tab:lattices}}
\end{table}

\section{Charm Results}

\begin{figure}[tbp]
  \centering
  \includegraphics[width=0.49\textwidth]{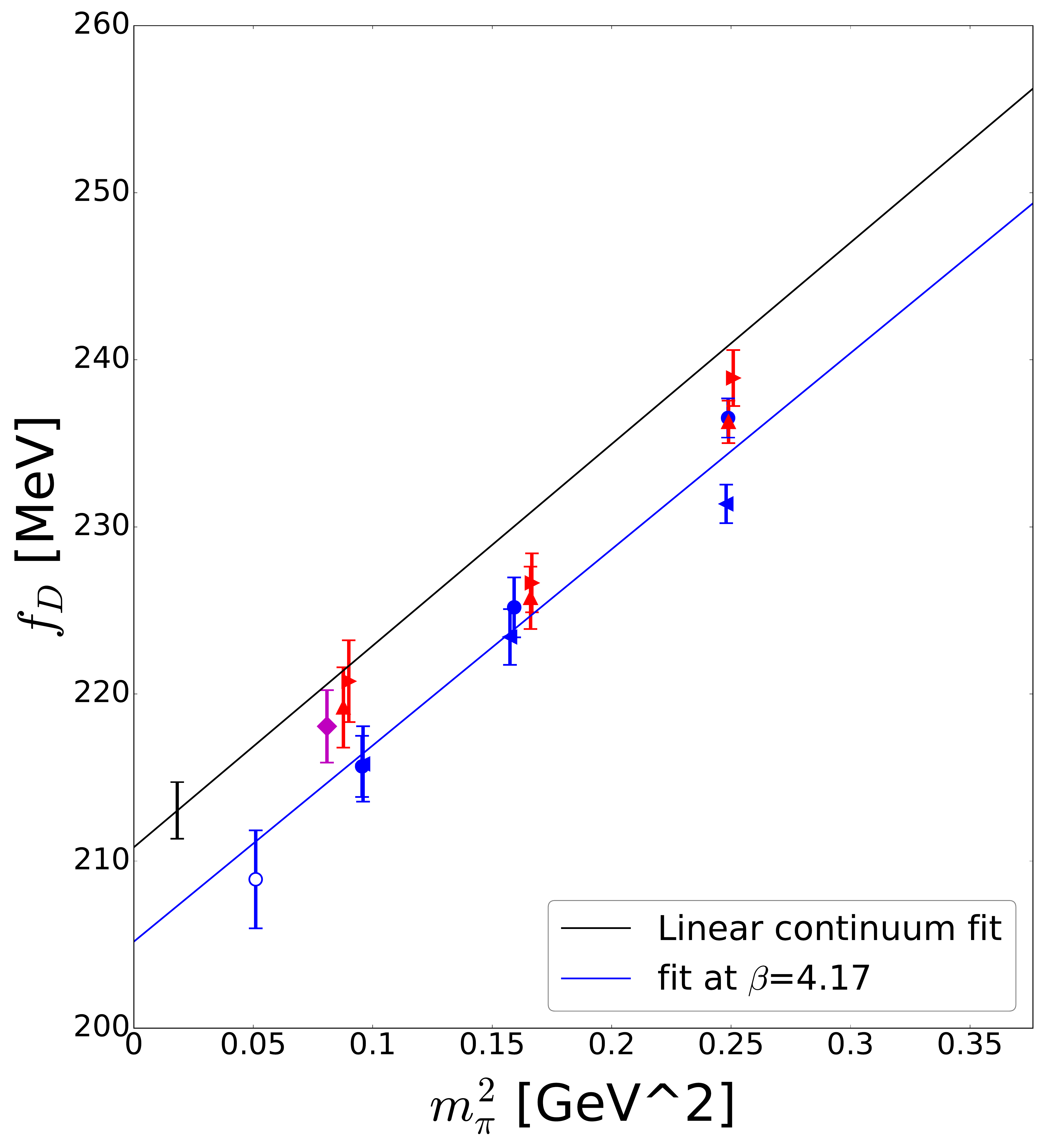}
  \includegraphics[width=0.49\textwidth]{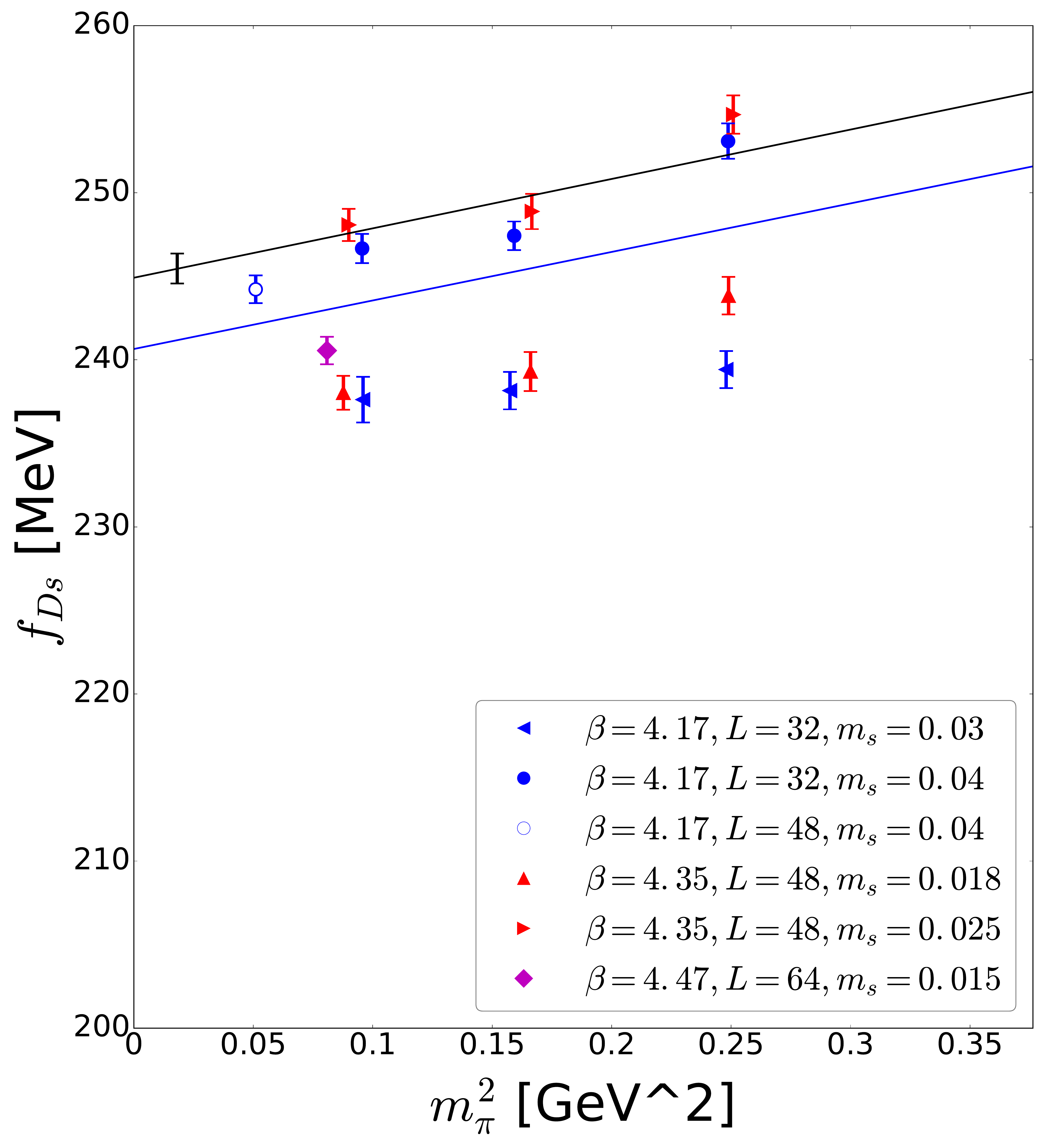}
  \caption{Plots of $f_D$ (left) and $f_{D_s}$ (right) vs
    $m_\pi^2$. The fit lines correspond to the continuum limit (black)
    and the coarsest lattice (blue). They are linear in $m_\pi^2$ and
    are interpolated to the physical strange point using
    $2m_K^2 - m_\pi^2$.}
  \label{fig:fdandfds}
\end{figure}

The first goal was to determine the decay constants $f_D$ and
$f_{D_s}$. These were determined from pseudo-scalar correlators at the
charm mass determined from a previous study~\cite{Fahycharm} with an
input of the spin averaged $c\bar{c}$ mass. The values for
$f_{D_{(s)}}$ for all of our ensembles can be seen in Figure
\ref{fig:fdandfds}. We preformed a global fit to all the ensembles
assuming linear dependence on the light and strange quark masses and
on the lattice spacing squared. The plots show lines corresponding to
the continuum limit (black) as well as the line evaluated at a lattice
spacing corresponding to our coarsest lattices. The difference between
the continuum limit and our coarsest lattice is roughly $2\%$. The
plots for $f_{D_s}$ (right panel) have fit lines which do not go
through the cluster of points because the data are simulated at
strange quark masses which sandwich the physical value.

The results of the global fit evaluated at the physical point are
\begin{align}
  \label{eq:fdresults}
  f_D&=212.8\pm 1.7 \pm 3.6 \, \text{MeV}, \\
  f_{D_s} &= 244.0 \pm 0.84 \pm 4.1\, \text{MeV},
\end{align}
with the errors being the statistical error and the
systematic error from the scale determination, respectively.

\section{Above Charm}

Since the cutoff effects at the charm mass were small we expect that
the results at slightly heavier quark masses are also under
control. We produced heavy-light and heavy-strange correlators using a
set of heavy masses above the charm mass. These were chosen to be
above the charm mass by factors of $1.25$ producing a sequence
$m_n = (1.25)^n m_c$. The bare charm mass is determined in a separate
study~\cite{Fahycharm}. The values for $m_n$ used on different $\beta$
values are shown in Table~\ref{tab:heavyquarkmasses}. For the finer
lattices we produced five bare quark masses above the charm mass,
while we limited to only three masses above charm for our coarsest
lattice as the bare quark mass $(1.25)^4m_c$ exceeds $1.0$. Note that
the maximum possible mass for the domain wall fermion is given by the
Pauli-Villars mass, which is $1$ in the standard implementation.

\begin{figure}[tbp]
  \centering
  \includegraphics[width=0.49\textwidth]{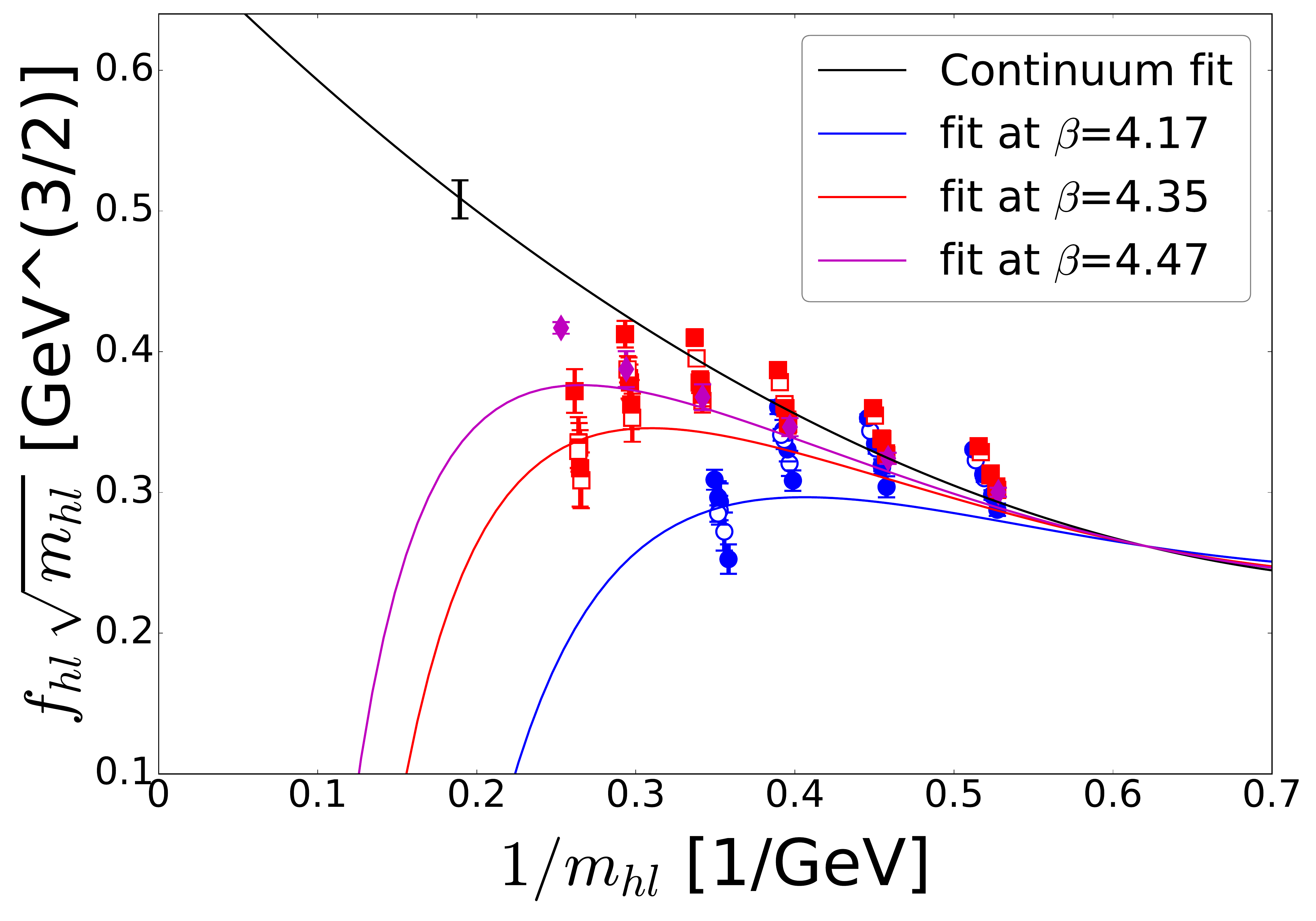}
  \includegraphics[width=0.49\textwidth]{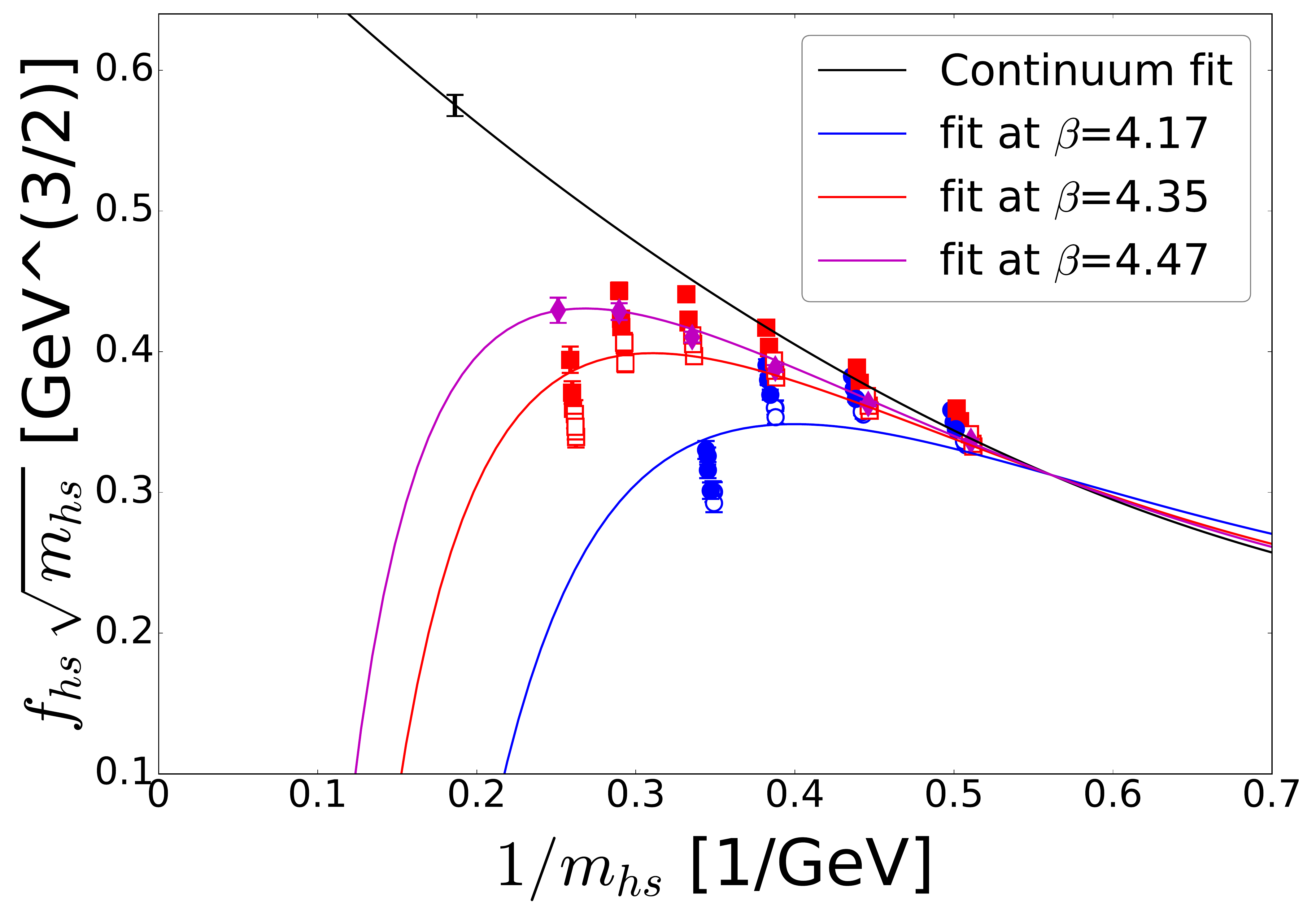}
  \caption{Values for $f\sqrt{m}$ for heavy-light (left) and heavy-strange (right) on all of our ensembles for each of the different heavy-quark masses from Table~\protect\ref{tab:heavyquarkmasses}. Data points are those of $\beta=4.17$ (blue), $4.35$ (red) and $4.47$ (magenta). Each horizontal cluster of points shows values for a particular choice of heavy quark mass with the value at the charm mass on the far right and increasing heavy quark masses going to the left. The black line indicates a global fit in physical limit as discussed in the text. The colored lines indicate the same fit parameters evaluated at the finite lattice spacings corresponding to our three choices of $\beta$. The empty points with error bars show the statistical uncertainly of the fit at the physical values for $m_B$ and \ensuremath{m_{B_s}}. \label{fig:heavyraw}}
\end{figure}

To see how well things scaled from the charm towards the bottom we
looked at the matrix element $f_{hx} \sqrt{m_{hx}}$ where $f$ is the
pseudo-scalar decay constant, $m$ is the pseudo-scalar meson mass, and
$x$ is either a strange $(s)$ quark or light $(\ell)$ quark . This was
chosen because in the heavy quark limit, $m_h \to \infty$, the
combination $f_{hx} \sqrt{m_{hx}}$ is known to scale as a
constant up to the anomalous dimension contribution (see below). Plots of this verses the inverse meson mass, for heavy-light
(left) or heavy-strange (right), are shown in
Figure~\ref{fig:heavyraw}. The results on all ensembles are plotted
together with the cluster of points on the right side being the values
at the charm mass and moving to the left each cluster being the next
choice for the heavy quark mass. It is clear that on the coarsest
lattice (blue) discretization effects become significant already at
$(1.25)^2 m_c$ and heavier. In the next section we attempt to account
for the leading order $a^2$ dependence. Here as a first attempt we
simply perform a global fit to all of the data including a term to
account for $a^2$ as well as $a^2m^2$ effects.

\begin{table}[tbp]
  \centering
  \begin{tabular}[h]{|l|r|r|r|r|r|r|}\hline
    Beta & $m_0 = m_c$ & $m_1$ & $m_2$ & $m_3$ & $m_4$ & $m_5$ \\ \hline
    4.17 & 0.4404 & 0.5505 & 0.6881 & 0.8600 &  & \\
    4.35 & 0.2729 & 0.3411 & 0.4264 & 0.5330 & 0.6661 & 0.8327 \\
    4.45 & 0.2105 & 0.2631 & 0.3289 & 0.4111 & 0.5139 & 0.6423 \\ \hline
  \end{tabular}
  \caption{Bare heavy quark masses where $m_0$ is the charm quark mass
    $m_n = \lambda^n m_0$. \label{tab:heavyquarkmasses}}
\end{table}

The global fit is performed using an ansatz
\begin{align}
  \label{eq:fit}
  \begin{split}
  &f_{hx}\sqrt{m_{hx}} = \p{\Phi_{\phys}} \p{1 + \displaystyle \frac{C_1}{m_h} + \frac{C_2}{m_h^2}} \\
  &\Phi_{\phys} = \p{1+\gamma_S \p{2M_K^2 - M_\pi^2} +\gamma_P\p{M_\pi^2-\p{M_\pi^\text{phys}}^2}+\gamma_A a^2+\gamma_{MA} (ma)^2}.
\end{split}
\end{align}
The basic assumption is the dependence on the inverse meson mass is a
polynomial preserving a constant in the limit of $m \to \infty$. The
$\Phi_{\phys}$ accounts for the extrapolation to the physical pion
mass, interpolation to the physical strange quark mass using
$2M_K^2 - M_\pi^2$, and extrapolation to the cotinuum limit with both
$m^2a^2$ and $a^2$ terms. Here the fit excludes the heaviest points on
the $\beta=4.17$ and $\beta=4.35$ lattices which have a bare quark
mass above $0.8$ where the discretization effect of order $(ma)^4$ and
higher are expected to be more significant. The fit curves
corresponding to each $\beta$ value and physical light/strange masses
are shown in Figure~\ref{fig:heavyraw}, We observe that the data
points drift away from the continuum curve (black) as $ma$ gets
large. In particular the heaviest points at $\beta=4.17$ and $4.35$
suffer from strong discretization effects.

\section{HQET Corrections}
\label{sec:hqet}

To understand the leading order cutoff effects we use ideas from Heavy
Quark Effective Theory (HQET). We closely follow the discussion of
\cite{LEPAGE199245,PhysRevD.55.3933}, which was mainly applied to
the Wilson-type fermions.

Expanding the energy of a free quark for low momentum we obtain
$ {E \approx m_1 + \frac{p^2}{2m_2} + \ldots }$, where on the lattice
the ``rest mass'', $m_1$, may not be equal to the ``kinetic mass'',
$m_2$. These corrections were computed years ago for Wilson
fermions~\cite{LEPAGE199245} which give simple corrections that have
been used to design actions suited for heavy
quarks~\cite{PhysRevD.55.3933}.  In the case of domain wall fermions
the expressions for these corrections are not as simple.

Starting with the propagator for domain-wall
fermions~\cite{PhysRevD.59.094505} and expanding in low momentum we
obtain $m_1$ and $m_2$ at tree level as well as the wave-function
renormalization, $A_{KLM}^{DW}$. The expressions for these factors are,
\begin{eqnarray}
  \label{eq:hqetcorrections}
  m_1 &= \log\p{1-W_0+\sqrt{\p{1-W_0}^2 - 1}}, \\
  m_2 &= \sqrt{W_0^2 - 2 W_0}\p{\frac{Q+1-2W_0}{(Q+1)+(Q-1)(2W_0^2+W_0)}},\\
  A_{KLM}^{DW} &= \frac{2}{(1-m^2)\s{1+\sqrt{\frac{Q}{1+4W_0}}}},
\end{eqnarray}
where
\begin{align}
  Q &= \p{\frac{1+m^2}{1-m^2}}^2 \quad \quad \text{and} \quad \quad W_0 = \frac{1+Q}{2} - \frac{\sqrt{3Q+Q^2}}{2}.
\end{align}
According to \cite{LEPAGE199245,PhysRevD.55.3933} we can then
re-scale the heavy quark mass in favor of $m_2$ from $m_1$ by adding
$m_2-m_1$, since the kinetic mass controls the motion of the heavy
quark inside the meson. Similarly, we divide the amplitude by
$A^{DW}_{KLM}$ to eliminate the leading discretization effect for the heavy
quark.

These corrections, however, turned out to be insufficient to account for the
heavy quarks propagator at short distances. If we numerically
integrate the free propagator and divided by $A_{KLM}$, they agree at
large time separations but for small separation they
disagree. This deviation from a simple exponential behavior may be due
to the non-locality of domain-wall fermions, which becomes relatively
more significant at large quark masses. To fully capture such lattice
artifacts we numerically integrate the free propagator for each bare
heavy quark mass, and divide our correlators by these and multiply
back by the continuum result, so that the heavy quark propagator
coincides that of the continuum at least at tree level.

\begin{figure}[tbp]
  \centering
  \includegraphics[width=0.49\textwidth]{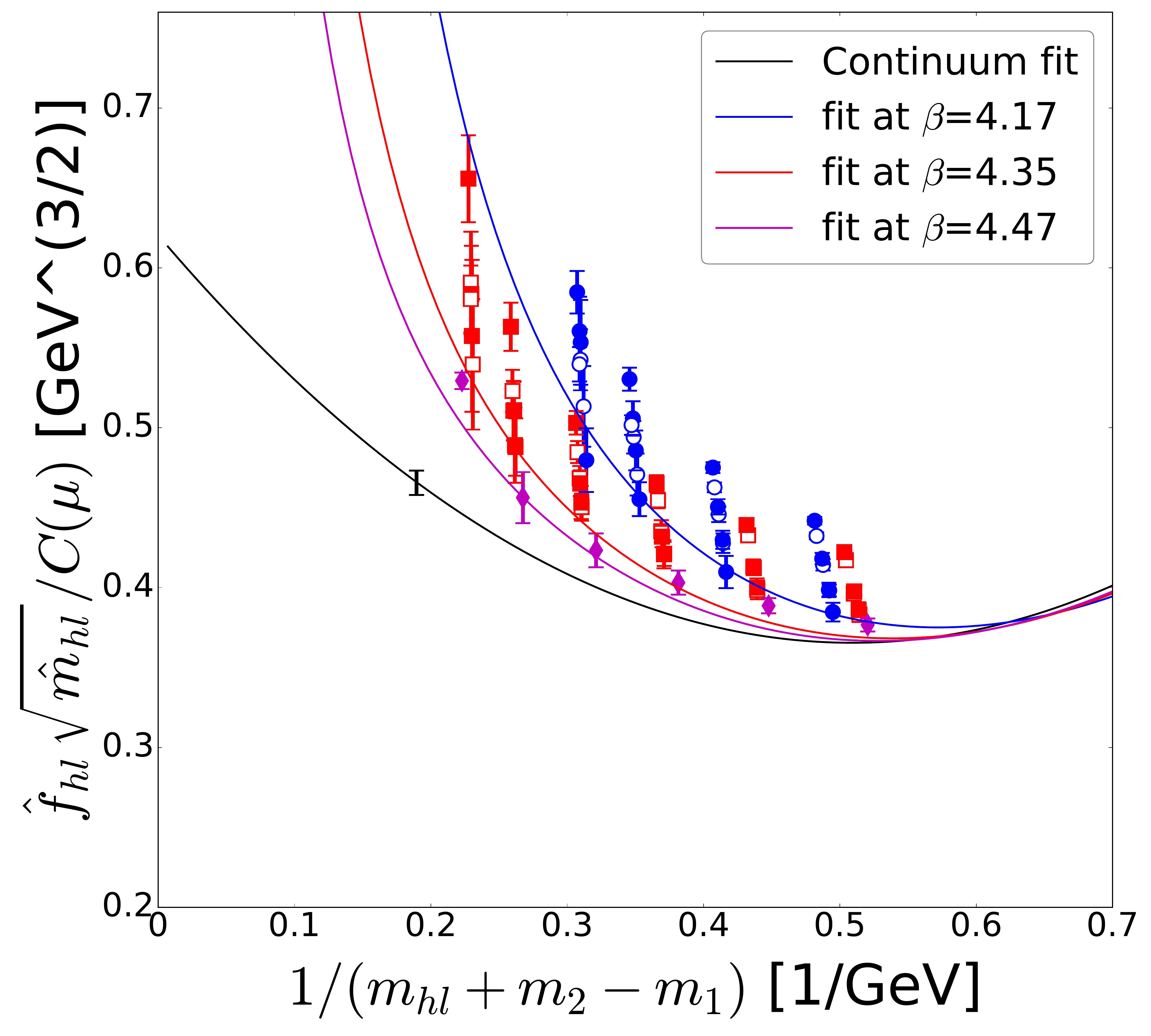}
  \includegraphics[width=0.49\textwidth]{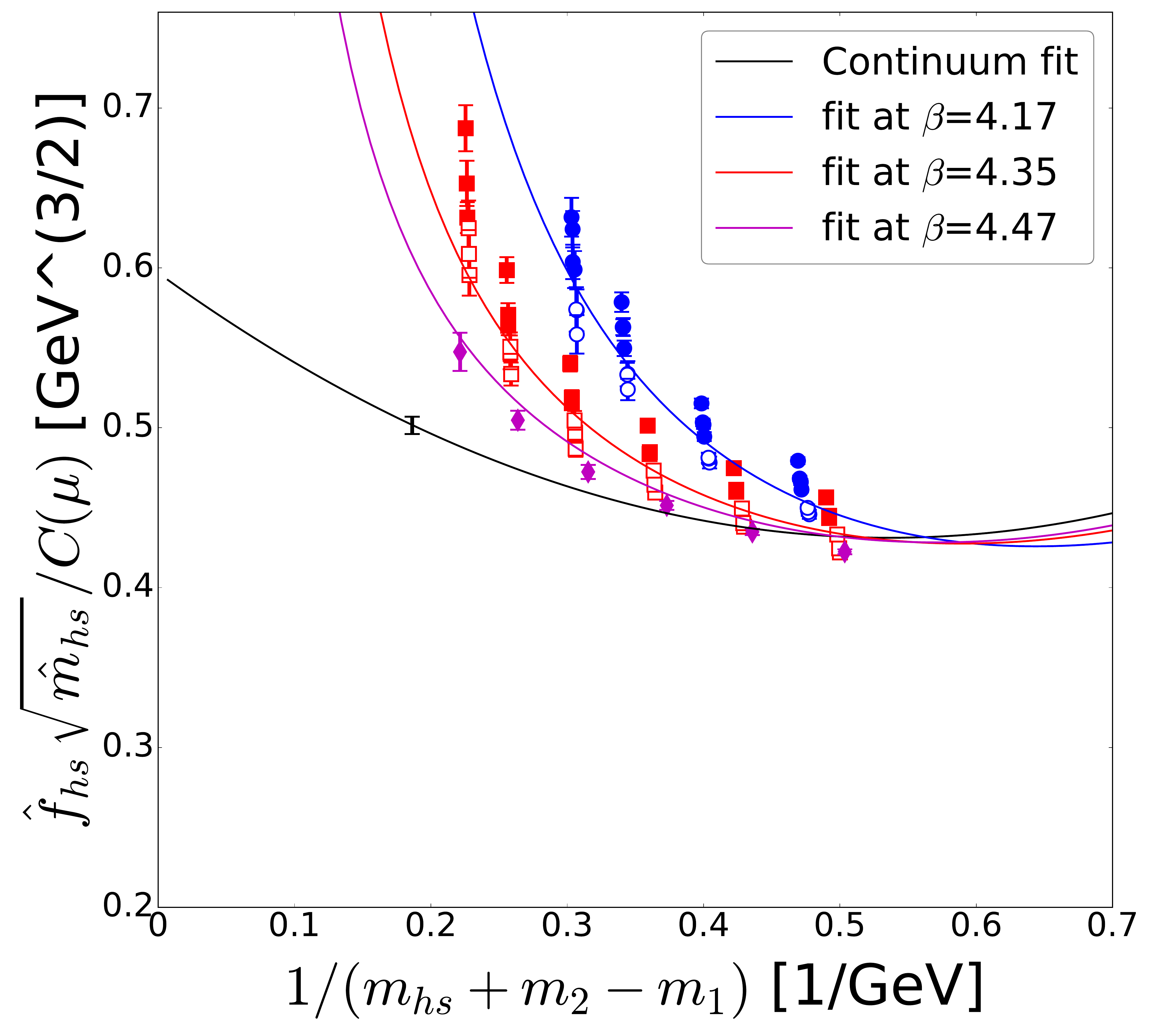}
  \caption{Same as Figure~\protect\ref{fig:heavyraw} with the corrections
    described in Section~\protect\ref{sec:hqet}. The anomalous dimension
    factors is also included. The fit now accounts for leading order
    $a^2m^2$ effects so the correction used is
    $\gamma_{MA}{{\alpha_s}} (ma)^2$. The black error bars on the fit
    give the statistical uncertainty at the $B$ and $B_s$ meson mass
    respectively.}
  \label{fig:heavycorrected}
\end{figure}

Applying these corrections we match the decay constants from QCD to
HQET with a factor accounting for the anomalous dimension $C(\mu)$. Perturbatve calculation of $C(\mu)$ is
available up to three loop, $\alpha_s^3$~\cite{Bekavac201046}.  These
results and fit are shown in Figure \ref{fig:heavycorrected}. These
results exhibit less divergence from the continuum limit of the values
for the heaviest quark masses indicating that we have successfully account
for the bulk of the leading $a^2m^2$ cutoff effects. The global fit to
the corrected data is performed in the same manner as before, see
(\ref{eq:fit}), but replacing the $m^2a^2$ term with
$\gamma_{MA} \alpha_s (ma)^2$ as we have accounted for the tree level
corrections.  The continuum limit (black line) as well as the fit for the
individual finite lattice spacings (colored lines) are plotted. The fit function
evaluated at the $B$ and $B_s$ mass yields,
\begin{align}
  f_B&=195.5 \pm 3.2 \pm 3.3 \text{ MeV} \\
  f_{B_s}&=218.2 \pm 1.9 \pm 3.7 \text{ MeV}.
\end{align}
These results are within $2\sigma$ of the current FLAG
average~\cite{FLAG2014}.

\section{Summary}
\label{sec:sum}

Using fine lattices we are able to obtain results for pseudo-scalar
decay constants above the charm mass at nearly the bottom
mass. Accounting for the leading discretization effects we are able to
extrapolate to the bottom quark mass and predict $f_B$ and
$f_{B_s}$. This is all done without requiring a specialized action for
heavy quarks as they were treated with the same domain wall action
used for light quarks.\\
\vspace{1cm}

Numerical simulations are performed on Hitachi SR16000 and IBM System
Blue Gene Solution at KEK under the support of its Large Scale
Simulation Program (No. 16/17-14). This research is supported in part
by the Grant-in-Aid of the MEXT (No. 26247043, 25800147) and by MEXT as
``Priority Issue on Post-K computer'' (Elucidation of the Fundamental
Laws and Evolution of the Universe) and JICFuS.

\bibliography{lattice_2016_fahy}






\end{document}